\documentclass[journal,comsoc,10pt]{IEEEtran}

\usepackage[T1]{fontenc}

\usepackage{amsmath,graphicx,mathtools,cite,soul,xcolor}
\allowdisplaybreaks

\normalsize

\newcommand{\bs}{\boldsymbol}

\begin{document}
\title{Optimized Cell Planning for Network Slicing in Heterogeneous Wireless Communication Networks }

\author{Florian Bahlke, Oscar D.~Ramos-Cantor, Steffen Henneberger and Marius Pesavento}

\markboth{IEEE Communications Letters}%
{Bahlke \MakeLowercase{\textit{et al.}}: Resource Allocation Planning For Network Slicing in Heterogenenous Wireless Communication Systems}

\maketitle

\begin{abstract}
We propose a cell planning scheme to maximize the resource efficiency of a wireless communication network while considering quality-of-service requirements imposed by different mobile services. In dense and heterogeneous cellular 5G networks, the available time-frequency resources are orthogonally partitioned among different slices, which are serviced by the cells. The proposed scheme jointly optimizes the resource distribution between network slices, the allocation of cells to operate on different slices, and the allocation of users to cells. Since the original problem formulation is computationally intractable, we propose a convex inner approximation. Simulations show that the proposed approach optimizes the resource efficiency and enables a service-centric network design paradigm.
\end{abstract}


\IEEEpeerreviewmaketitle

\section{Introduction} \label{sec:intro}
\IEEEPARstart{I}{n} recent years, heterogeneous architectures with dense cell deployments have been identified as a promising technology candidate to fulfill increasing performance- and service requirements of future wireless communication networks. The fundamental advantage of such networks is that they complement existing infrastructure and technology, and therefore require lower levels of commitment and economic risk from network operators \cite{andrews14}. As a key feature, future wireless networks are envisioned to operate based on a "Network Slice Layer" and a "Service Layer" \cite{NGMN_slicing,foukas17}. The function of a network slice is to aggregate sets of network resources from the underlying physical layer to provide specific services in the service layer \cite{ferrus18}. \\
From a network operators perspective one of the greatest concerns is how to enable the coexistence of a a variety of services, with very diverse requirements regarding reliability, data throughput or latency, in a network that due to its dense cell deployment becomes increasingly interference-limited \cite{andrews16}. To achieve the high signal-to-interference-plus-noise-ratio (SINR) levels that are necessary for high-reliability or high-throughput slices, the network can only be "densified" to a limited extent if co-channel operation is being considered. To mitigate the significant interferences caused by network densification, multiple slices operating on orthogonal sets of physical time-frequency resources might be necessary \cite{andrews16,NGMN_slicing}. This approach originates from conventional cell planning, but can be adapted to optimize resource allocation to slices providing different services. Dedicated network optimization methods for dense heterogeneous slicing networks are still scarce in current literature \cite{andrews16}. \\
In this work, we propose an approach for network time-frequency resource planning based on maximizing the resource efficiency of the network by joint optimization of the resource assignment to network slices, the allocation of slices to different operating cells, and the allocation of users to cells. We also demonstrate how SINR-requirements and bandwidth efficiencies of the transmission schemes dedicated to different services can be incorporated into the network optimization process. \\

\section{System Model} \label{sec:sys}
We consider a wireless communication network with $K$ cells and $M$ demand points (DP), indicated by $k=1,\ldots,K$ and $m=1,\ldots,M$, respectively, and assume each cell to be defined as the coverage area of one base station. As typical for the current fourth and upcoming fifth generation networks, we assume a heterogeneous cellular architecture with macro-cells (MC) and small cells (SC). The data demand in terms of a requested rate of DP $m$, which may represent the demand of single user nodes or accumulated demands from hotspots, is denoted as $D_m$. Bandwidth resources available to the network are divided among $I$ slices, indicated by $i=1,\ldots,I$. These slices may be designed to provide different services, with varying minimum SINR requirements $\gamma_{i}^\text{MIN}$ and bandwidth efficiencies $\eta_{i}^\mathrm{I}$, which represent the ratio of the bandwidth available for data transmission to the total available bandwidth. Other service requirements may include peak data rates and latency, which both can be optimized in higher network layers. A low-latency transmission scheme for example would rely on very small packet sizes, which can be modeled with a decreased bandwidth efficiency in the proposed scheme \cite{majewski10}. The information in which of the slices $i$ a DP $m$ can be served is specified by parameter $S_{im}$, where $S_{im}=1$ if DP $m$ can be served in slice $i$, and $S_{im}=0$ otherwise. In this work, we assume without loss of generality that each cell operates in a single slice. A cell that operates in multiple slices can be modeled as multiple "virtual" cells in the same location. The allocation of cells to slices is indicated in matrix $\bs{B} \in \{0,1\}^{I \times K}$, where $B_{ik}=1$ if cell $k$ operates in slice $i$ and $B_{ik}=0$ otherwise. The SINR $\gamma_{km}$ of DP $m$ served by the base station in cell $k$ is formulated as
\begin{equation} \label{eq:SINR_def}
\gamma_{km} = \frac{p_k g_{km}}{\sum_{j\setminus{\{k\}}}^K \sum_{i=1}^I B_{ik} B_{ij} p_j g_{jm} + \epsilon}
\end{equation}
where $p_k$ is the power spectral density of the transmitted signal of cell $k$, $g_{km}$ is the combined attenuation factor from cell $k$ to user $m$ resulting from antenna gains and path loss and $\epsilon$ is the power spectral density of additive white Gaussian noise. The sum over $j\setminus{\{k\}}$ refers to the set of all cells $j=1,\ldots,K$ except for $j=k$. The model in 1 corresponds to an OFDMA system with full frequency reuse between cells \cite{majewski10}. Note that the term $\sum_{j\setminus{\{k\}}}^K \sum_{i=1}^I B_{ik} B_{ij} p_j g_{jm}$ indicates the interference from those cells $j$ which are serving, and therefore interfering, in the same slice $i$ as cell $k$. We indicate the allocation of DPs to cells with the binary matrix $\bs{A} \in \{0,1\}^{K \times M}$, with its elements $A_{km}=1$ if DP $m$ is allocated to cell $k$, and $A_{km}=0$ otherwise. The bandwidth efficiency modifying factor related to the type of cell $k$ (e.g.~MC or SC) used for the transmission is indicated as $\eta_{k}^\mathrm{K}$. Based on the cell load computation outlined in \cite{majewski10}, cell $k$ is not overloaded if the following condition is satisfied:
\begin{equation} \label{eq:load_def}
\sum_{m=1}^M A_{km} D_m f\left(\gamma_{km} \right) \leq \eta_{k}^\mathrm{K} \sum_{i=1}^I \left(B_{ik} \eta_{i}^\mathrm{I} w_i \right)  \; \forall k,
\end{equation}
where $f(\gamma) = 1/\log_2(1+\gamma)$, $w_i$ is the bandwidth allocated to slice $i$. We denote the total system bandwidth as $\overline{w}$, such that for any viable network configuration $\sum_{i=1}^I w_i \leq \overline{w}$.

\section{Resource Allocation Planning} \label{sec:method}
To maximize the overall resource efficiency of the network, we aim to find a cell- and slicing- configuration such that the requests from all DP are fulfilled with a minimum amount of bandwidth resources used. This problem is equivalent to maximizing the amount of "unused" resources $Z = \overline{w} - \sum_{i=1}^I w_i$. The practical use of this approach is that the unused spectral resources, after optimizing the network using our proposed approach, can be utilized for example to further improve the data rates of selected users. For this purpose a mixed-integer nonlinear optimization problem (MINLP) can be formulated as follows:
\begin{subequations} \label{eq:orig}
	\begin{align} 
	\underset{Z,\bs{w} \!,\bs{A},\bs{B}}{\mathrm{max.}} \: & Z  \label{eq:orig_obj} \\ 
	\mathrm{s. \ t.} \ & \sum_{i=1}^I w_i + Z = \overline{w} \label{eq:orig_w}\\
	& \sum_{k=1}^K A_{km} = 1 \; \forall m \label{eq:orig_A} \\
	& \sum_{i=1}^I B_{ik} \leq \min \left\{1, \sum_{m=1}^M A_{km} \right\} \; \forall k \label{eq:orig_B1}\\
	& \sum_{k=1}^K A_{km} B_{ik} \leq S_{im} \; \forall i,m \label{eq:orig_S} \\
	&\gamma_{km} =  \frac{p_k g_{km}}{\sum_{j\setminus{\{k\}}}^K \sum_{i=1}^I B_{ik} B_{ij} p_j g_{jm} + \epsilon} \; \forall k,m \label{eq:orig_gammadef} \\ 
	& A_{km} \gamma_{km} \geq A_{km} \gamma_{i}^\text{MIN} \ \forall k,\left\{m,i | \sum_k A_{km} B_{ik}=1 \right\} \label{eq:orig_gammacon}\\
	& \sum_{m=1}^M A_{km} D_m f\left(\gamma_{km} \right) \leq \eta_{k}^\mathrm{K} \sum_{i=1}^I \left(B_{ik} \eta_{i}^\mathrm{I} w_i \right)  \; \forall k \label{eq:orig_load}\\
	& Z,w_{i} \in \mathbb{R}_{0+} \; \forall i \label{eq:orig_reals} \\
	& A_{km},B_{ik} \in \{0,1\} \; \forall i,k,m \label{eq:orig_bins}
	\end{align}
\end{subequations}
In problem \eqref{eq:orig}, equality \eqref{eq:orig_A} forces each DP to be allocated to exactly one cell, while inequality \eqref{eq:orig_S} allows it to be allocated only to its requested slice(s), as specified by $S_{im}$. Constraints \eqref{eq:orig_B1} cause each cell to operate in at most one slice, and only if it has users allocated to it. With \eqref{eq:orig_gammacon}, an allocated DP-cell connection needs to fulfill the SINR requirement of the slice requested by the DP. \\
Problem \eqref{eq:orig} is a nonconvex mixed-integer program. Especially the dependency of the interference-plus-noise term in Eq.~\eqref{eq:orig_gammadef} on $\bs{B}$ and multiple bilinear terms in other constraints render the problem computationally intractable to solve. In the following we introduce an inner approximation of the original problem to obtain a mixed integer linear program (MILP), for which computationally efficient general purpose solvers are available.  \\
This inner approximation is performed in three steps: firstly, the interference for each connection is upper bounded with a set of discrete interference scenarios, secondly the SINR- and load-computation are reformulated with said approximation, and finally all bilinear products of optimization parameters are replaced with equivalent linear formulations using a lifting procedure.\\
To provide an approximation of the SINR expression as defined in Eq.~\eqref{eq:orig_gammadef}, we introduce a set of discrete interference levels $\Psi_{nkm}$, indicated by $n=1,\ldots,N$, which are precomputed for each pair of $(k,m)$, such that they represent the most relevant low- to medium-SINR scenarios. To ensure the feasibility of the approximate problem, the full interference scenario, i.e.~$\Psi_{1km}=\sum_{j \setminus \{k\}} p_j g_{jm} + \epsilon$ is considered as one of the relevant interference scenarios. The second relevant interference level is the scenario where the strongest interfering cell is inactive or operating in another slice, for example $\Psi_{2km}=\Psi_{1km} - \max_{j\setminus \{k\}} p_j g_{jm}$, . Usually the removal of the first- and second-strongest interfering cell has the highest impact on achievable rates. The algorithm is then designed in such a way that the discrete interference level used as an approximation is always an over-estimator of the actual interference plus noise. This is achieved by adding in the reformulated problem the following constraint:
\begin{equation} \label{eq:intless}
\sum_{j\setminus{\{k\}}}^K \sum_{i=1}^I B_{ik} B_{ij} p_j g_{jm} + \epsilon \leq \sum_{n=1}^N \Theta_{nkm} \Psi_{nkm}  \; \forall k,m
\end{equation}
where we denote $\Theta_{nkm}=1$ if interference scenario $n$ is used as an approximation for the link between DP $m$ and cell $k$, and $\Theta_{nkm}=0$ otherwise. Since exactly one interference scenario applies for each pair of cell $k$ and DP $m$, it has to hold that $\sum_n \Theta_{nkm} =1 \; \forall k,m$. The elements $\Theta_{nkm}$ are arranged in the three-dimensional binary array $\bs{\Theta} \in  \{0,1\}^{N \times K \times M}$. The proposed inner approximation of Eq.~\eqref{eq:orig_load} is the following:
\begin{equation} \label{eq:loadlong}
\sum_{m=1}^M \sum_{n=1}^N D_m \Theta_{nkm} A_{km} f\left( \frac{p_k g_{km}}{\Psi_{nkm}} \right) \leq \eta_{k}^\mathrm{K} \sum_{i=1}^I \left( \eta_{i}^\mathrm{I} B_{ik} w_i \right) \; \forall k \\
\end{equation}
The term $f\left(p_k g_{km}/\Psi_{nkm} \right)$ in Eq.~\eqref{eq:loadlong} can be pre-computed for all combinations of $(n,k,m)$. Approximating the constraints \eqref{eq:orig_load} with \eqref{eq:intless} and \eqref{eq:loadlong} leads to an upper bound approximation of the interference and accordingly the required bandwidth for each user. Solutions obtained from using the latter set of constraints are therefore feasible for the original problem. We approximate the constraints \eqref{eq:orig_gammacon} correspondingly with
\begin{equation} \label{eq:gammacon_full}
A_{km} p_k g_{km} \geq \gamma_{i}^\text{MIN} \sum_{n=1}^N \Theta_{nkm} A_{km} \Psi_{nkm} \; \forall k,\{m,i|S_{im}=1\}.
\end{equation}
The resulting optimization problem is still nonlinear because of multiple bilinear products of two optimization variables in the constraints, specifically $A_{km} B_{ik}$ in Eq.~\eqref{eq:orig_gammacon} and Eq.~\eqref{eq:orig_S}, $\Theta_{nkm} A_{km}$ in Eq.~\eqref{eq:gammacon_full} and Eq.~\eqref{eq:loadlong}, $B_{ik} B_{ij}$ in Eq.~\eqref{eq:intless} and $B_{ik} w_i$ in Eq.~\eqref{eq:loadlong}. To linearize the corresponding constraints, we define the constraint sets $\mathcal{L} \coloneqq \{ \left(r,\overline{r},b,a \right) \in \mathbb{R}_{0+} \times \mathbb{R}_{0+} \times \{0,1\} \times \mathbb{R}_{0+}: a\geq r-(1-b)\overline{r}, a\leq r,a\leq b \overline{r} \}$ and $\mathcal{B} \coloneqq \{ \left(b_1,b_2,c\right) \in \{0,1\} \times \{0,1\} \times \{0,1\}: c \leq b_1, c\leq b_2, c \geq b_1+b_2-1 \}$.
The constraint set $\mathcal{L}$ represents a linearization of $rb=a$ if $\left(r,\overline{r},b,a \right) \in \mathcal{L}$ with a binary $b$ and $\overline{r}$ being an upper bound on the real parameter $r$. Similarly, the constraint set $\mathcal{B}$ enforces $b_1 b_2 = c$ for two binary parameters $b_1$ and $b_2$ if $\left(b_1,b_2,c\right) \in \mathcal{B}$. \\
To replace the aforementioned bilinear products, we use a lifting strategy that linearizes the problem at the cost of increased dimensionality. We introduce the auxiliary optimization parameters $X_{ikm} \triangleq A_{km} B_{ik}$, $\Phi_{nkm} \triangleq \Theta_{nkm} A_{km}$, $\Lambda_{ikj} \triangleq B_{ik} B_{ij}$ and $\Omega_{ik} \triangleq B_{ik} w_i$, with their elements arranged in the three-dimensional binary arrays $\bs{X} \in  \{0,1\}^{I \times K \times M}$, $\bs{\Phi} \in  \{0,1\}^{N \times K \times M}$,  $\bs{\Lambda} \in  \{0,1\}^{I \times K \times K}$ and the matrix $\bs{\Omega} \in  \mathbb{R}_{0+}^{I \times K}$, respectively. To ensure that these auxiliary parameters are equal to the bilinear products of the original optimization parameters the linear inequality constraints in $\mathcal{L}$ and $\mathcal{B}$ are introduced. This results in the following MILP:
\begin{subequations} \label{eq:linprob}
	\begin{flalign}
		\underset{Z,\bs{w},\bs{A},\bs{B},\bs{X},\bs{\Lambda},\bs{\Theta},\bs{\Phi},\bs{\Omega}}{\mathrm{maximize}} \quad  Z &&
	\end{flalign}
	\begin{flalign}
		\mathrm{s.t.} \; & \eqref{eq:orig_w}-\eqref{eq:orig_B1} \nonumber \\
		& \sum_{k=1}^K X_{ikm} \leq S_{im} \; \forall i,m \\
		& \gamma_{i}^\text{MIN} \left( \sum_{n=1}^N \Phi_{nkm} \Psi_{nkm} - \xi \left(1-\sum_{k=1}^K X_{ikm} \right) \right) \nonumber \\
		& \quad \leq A_{km} p_k g_{km}  \forall k,i,m  \label{eq:linprob_gammacon} \\
		& \sum_{m=1}^M \sum_{n=1}^N D_m \Phi_{nkm} f\left( \frac{p_k g_{km}}{\Psi_{nkm}} \right) \leq \eta_{k}^\mathrm{K} \sum_{i=1}^I \left( \eta_{i}^\mathrm{I} \Omega_{ik} \right) \; \forall k \\
		& \sum_{i=1}^I \sum_{j=1}^K \Lambda_{ikj} p_j g_{jm} + \epsilon \leq \sum_{n=1}^N \Theta_{nkm} \Psi_{nkm}  \; \forall k,m  \\
		& \sum_n \Theta_{nkm} =1 \; \forall k,m \\
		& \{ w_i, \overline{w},B_{ik}, \Omega_{ik} \} \; \in \; \mathcal{L} \; \forall i,k \\
		& \{ \Theta_{nkm},A_{km},\Phi_{nkm} \} \; \in \; \mathcal{B} \; \forall n,k,m \\
		& \{ A_{km},B_{ik},X_{ikm} \} \; \in \; \mathcal{B} \; \forall i,k,m \\
		& \{ B_{ik}, B_{ij}, \Lambda_{ikj} \} \; \in \; \mathcal{B} \; \forall i, j \neq k; \Lambda_{ikj}=0 \; \forall i,j=k \\
		& Z,w_{i},\Omega_{ik} \in \mathbb{R}_{0+} \; \forall i,k \\
		& A_{km},B_{ik},\Theta_{nkm},\Phi_{nkm},\Lambda_{ikj},X_{ikm} \in \{0,1\} \forall i,k,j,m,n
	\end{flalign}
\end{subequations}
In problem \eqref{eq:linprob}, Eq.~\eqref{eq:linprob_gammacon} is a big-M reformulation \cite{cheng13} where the right-hand side term $\xi (1-\sum_{k=1}^K X_{ikm})$ in this inequality ensures that the SINR-constraint is always fulfilled if $\sum_{k=1}^K X_{ikm} = 0$, i.e.~when DP $m$ is not serviced in slice $i$. For this property to hold, the parameter $\xi$ needs to be chosen such that $\xi \geq \max_{n,k,m} \Psi_{nkm}$.\\
The problem formulated in \eqref{eq:linprob} is linear in all optimization variables and therefore can be solved using conventional MILP solvers.

\section{Simulation Results} \label{sec:sims}
\begin{figure}
	\centering	
	\includegraphics[width=0.47\textwidth]{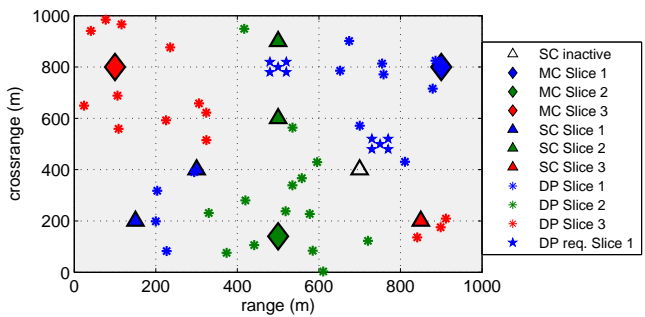}
	\caption{Network scenario, 3 macro- \& 6 small cells, 50 demand points}
	\label{fig:special}
\end{figure}

We simulate a mobile communication network as illustrated in Fig.~\ref{fig:special}, with 3 macro- and 6 pico cells distributed over an area of $1000$m $\times 1000$m. The pico cells are deployed close to the borders of the coverage areas of the macro cells. We model the macro cells with a transmit power spectral density of $-27$ dBm/Hz and $10$ dBi antenna gain, resulting in an equivalent effective isotropic radiated power (EIRP) of $-17$ dBm/Hz. Similarly, the pico cells are simulated with $-37$ dBm/Hz transmit power spectral density and $5$ dBi antenna gain, resulting in an EIRP of $-32$ dBm/Hz. These EIRP values are used for the parameters $p_k$. The path loss is simulated according to the specifications in 3GPP TS 36.814 Model 1 \cite[p.~61]{3GPP}, with additional 5dB log-normal shadow fading. The overall system bandwidth is $\overline{w} =20 \ \text{MHz}$. We pre-compute the following $N=3$ interference scenarios for $\Psi_{nkm}$ as introduced in Sec.~\ref{sec:method}: one where all cells are fully interfering, one where the strongest interfering cell is inactive or operating in a different slice, and one where the both first- and second-strongest interfering cells are inactive or operating in a different slice. The proposed optimization problem \eqref{eq:linprob} is solved using MATLAB with the CVX toolbox \cite{cvx} and Gurobi as a MILP solver \cite{gurobi}. The network scenarios described in this section have been solved on a standard workstation with an Intel i7-7600 processor, with a solver time of approximately one minute for each scenario for $M=40$ DPs. \\
An example network scenario is shown in Fig.~\ref{fig:special} to illustrate a typical result from the proposed method, where we added two clusters with five "special" DPs each that explicitly request service from a "high reliability" slice $i=1$ with a decreased bandwidth efficiency and higher SINR requirements. As observable, the macro cell in the upper right corner of the map is reserved almost exclusively for the service of these special DPs. \\
We analyze the effect of cell planning with multiple orthogonal resource pools on the resource efficiency of the system, as optimized by solving problem \eqref{eq:linprob}. As a baseline method to optimize the resource efficiency we use the state-of-the-art approach of full frequency reuse, and an allocation of DPs to the cell providing the strongest signal. To allow for a comparison with this approach, we model the slices in our proposed method with equal parameters, specifically $\gamma_{i}^\text{MIN} = -7\text{dB} \; \forall i$ and $\eta_{i}^\mathrm{I} = 1 \; \forall i$. 

\begin{figure}
	\centering	
	\includegraphics[width=0.47\textwidth]{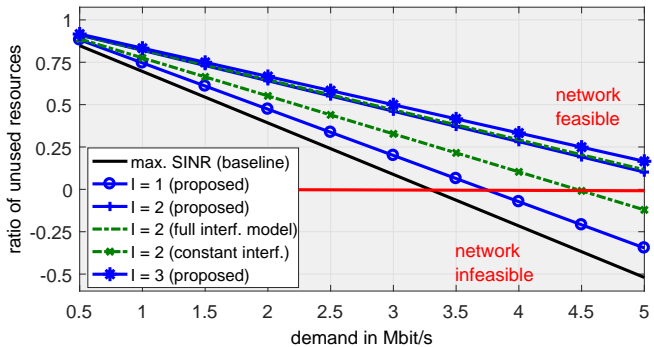}
	\caption{Ratio of unused resources $Z/\overline{w}$ for varying maximum number of slices $I$ over demand point demand $D_m$}
	\label{fig:rescomp}
\end{figure}

\begin{figure}
	\centering	
	\includegraphics[width=0.47\textwidth]{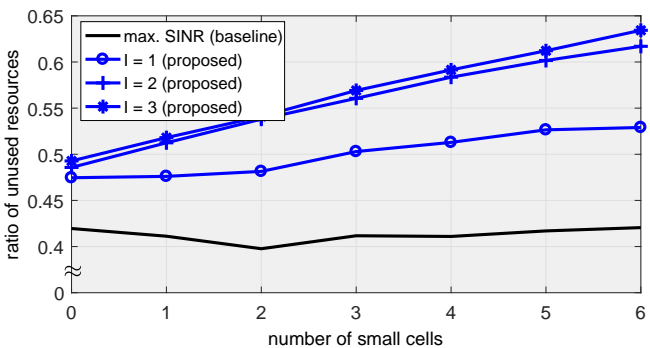}
	\caption{Ratio of unused resources $Z/\overline{w}$ for varying maximum number of slices $I$ over number of small cells}
	\label{fig:rescomp2}
\end{figure}

In a first simulation, we simulate 250 network scenarios with the DP demand in each scenario increasing from $D_m=0.5 \ \text{Mbit/s}$ to $D_m=5 \ \text{Mbit/s}$, and $M=40$ randomly placed DPs for each scenario, where we average the resulting levels of unused resources $Z$ over all scenarios. The network for this scenario is simplified from that shown in Fig.~\ref{fig:special} such that only the three macro cells and the small cell in the center are deployed. The DPs are randomly placed in the simulated area using a uniform probability distribution. As observable in Fig.~\ref{fig:rescomp}, the proposed method with $I=1$ yields a resource efficiency slightly higher than the baseline method, but remark that both methods require more resources than are actually available for high demand, which means that $Z$ is negative. For $I=2$, we show the remaining resources also for modified versions of the algorithm. In the first modified algorithm, indicated as "full interference model", we show that selecting $\Psi_{nkm}$ such that all possible combinations of active cells are considered only provides marginal performance increases. The second modified algorithm assumes a constant interference model used for example in \cite{caballero17} and the references therein, which shows significantly decreased resource efficiency due to the worse approximation of the actual interference. \\
In a second numerical experiment with the same parameters as that from Fig.~\ref{fig:rescomp}, we simulate 500 network scenarios with $M=40$ and $D_m$ of 2 Mbit/s, where the 6 small cells are one by one activated in a location randomly chosen from the 6 available locations shown in Fig.~\ref{fig:special}. This simulation is designed to evaluate the benefits of network densification, represented by the number of additionally deployed small cells. The results are shown in Fig.~\ref{fig:rescomp2}. As observable, the resource efficiency for the max.~SINR method overall shows no real benefit from densifying the network. The proposed method however already shows some gains for $I=1$, where only user allocation and cell activity status is optimized. If additionally $I=2$ or $I=3$ slices operating on orthogonal resource pools are allowed, the proposed method demonstrates gains in resource efficiency and the benefits of network densification.

\section{Conclusion} \label{sec:conc}
We introduced an optimization method for maximizing the resource efficiency of a wireless communication network by joint optimization of resource distribution to network slices, dimensioning of the slices, and allocation of users to cells. The proposed method can enable a service-centric organization of the network incorporating the bandwidth efficiencies and SINR-requirements of different transmission schemes in the network design process. Future research could be dedicated to finding heuristic methods that can provide decentralized optimization for larger networks.


\bibliographystyle{IEEEtran}
\bibliography{refs}

\end{document}